\documentclass[aps,prd,twocolumn,superscriptaddress,showpacs,bibnotes,preprintnumbers,amsmath,amssymb]{revtex4}
\usepackage{graphicx}% Include figure files
\usepackage{dcolumn}% Align table columns on decimal point
\usepackage{bm}% bold math
\usepackage{epstopdf}
\usepackage{mathtools}
\usepackage{natbib}
\usepackage{captcont}

\def\apj{Astrophys. J.}
\def\prd{Phys. Rev. D}
\def\mnras{Mon. Not. R. Astron. Soc.}
\def\jcap{J. Cos. Astropart. Phys.}
\def\aap{Astron. Astrophys.}

\def\apjl{Astrophys. J. Letter}

\begin{document}
\title{TeV-PeV neutrinos over the atmospheric background: originating from two groups of sources?}

\affiliation{Key Laboratory of Dark Matter and Space Astronomy, Purple Mountain Observatory, Chinese Academy of Sciences, Nanjing 210008, China }

\author{Hao-Ning He}
\affiliation{Key Laboratory of Dark Matter and Space Astronomy, Purple Mountain Observatory, Chinese Academy of Sciences, Nanjing 210008, China }
\author{Rui-Zhi Yang}
\affiliation{Key Laboratory of Dark Matter and Space Astronomy, Purple Mountain Observatory, Chinese Academy of Sciences, Nanjing 210008, China }
\author{Yi-Zhong Fan$^\ast$}
\affiliation{Key Laboratory of Dark Matter and Space Astronomy, Purple Mountain Observatory, Chinese Academy of Sciences, Nanjing 210008, China }
\author{Da-Ming Wei}
\affiliation{Key Laboratory of Dark Matter and Space Astronomy, Purple Mountain Observatory, Chinese Academy of Sciences, Nanjing 210008, China }

\begin{abstract}
In addition to the two $\sim 1$ PeV neutrinos,
the IceCube Collaboration recently reported a detection of 26 neutrino candidates at energies from $30$ TeV to $250$ TeV,
implying a confidence level of $4.3\sigma$ over the atmospheric background. 
We suggest that these TeV-PeV non-atmospheric neutrinos may originate from two groups of sources, 
motivated by the non-detection of neutrinos in the energy range 250 TeV$-$ 1 PeV in current data. 
If intrinsic, the non-detection of 250 TeV$-$1 PeV neutrinos disfavors the single power-law spectrum 
model for the TeV$-$PeV non-atmospheric neutrinos at a confidence level of $\sim 2\sigma$.
We then interpret the current neutrino data with a two-component spectrum model. One has a flat spectrum with a cutoff at the energy $\sim 250$ TeV and the other has a sharp peak at $\sim 1$ PeV. The former is likely via $pp$ collision while the latter may be generated by the photomeson interaction. \end{abstract}
\pacs{95.85.Ry, 95.85.Hp, 98.70.Sa}
\maketitle

The construction of IceCube was completed in 2010,
and data were collected from May 2010 to May 2011 with 79 strings
and an effective livetime of 285.8 days,
and from May 2011 to May 2012 with the full 86-string detector
and a livetime of 330.1 days.
In a search for very-high energy neutrinos, i.e., GZK neutrinos produced via ultra-high energy
cosmic rays interacting with cosmic microwave background (CMB) photons and extragalactic background light (EBL) photons
 \cite{Greisen1966, Zatsepin1966, Kotera2010},
the IceCube Collaboration discovered
two cascade events with energy $\sim 1$ PeV \cite{Ishihara2012,Aartsen2013},
which is $2.8\sigma$ beyond the atmospheric background.
The energies of these two events are the highest so far
but lower than that expected for GZK neutrinos.
However, if assuming an unbroken $E^{-2}$ power-law spectrum,
there should be about $8-9$ more events observed above $2$ PeV.

Quite recently, they carry on the high-energy contained vertex search
to analysis the 662 days data from May 2010 to May 2012,
which is sensitive at energy
above $50$ TeV, and at 1PeV, three times as sensitive as GZK search.
At the IceCube Particle Astrophysics Symposium (IPA-2013),
they reported an observation of 28 neutrino candidates,
including 26 events with energy from $30$ TeV to $250$ TeV,
and the two $\sim 1$ PeV neutrinos detected previously,
implying a significance level of $4.3\sigma$
over the atmospheric background of $10.6^{+4.5}_{-3.9}$ events \cite{Whitehorn2013}.

The non-detection of events above $\sim 2$ PeV and the excess of events over atmospheric background
is the two momentous features of the new IceCube data.
In favor of the astrophysical origins,
the non-detection above $\sim 2$ PeV may indicate a broken or peaked spectrum with a break/cutoff or a peak
in the PeV band \cite{Whitehorn2013, Murase2013},
or a soft unbroken spectrum with index larger than 2.3 \cite{Anchordoqui2013,Laha2013}.
He et al. \cite{He2013} calculated the neutrino spectrum from pp collision in ultra-luminous
infrared galaxies (ULIRGs), and suggested a cutoff of neutrinos spectrum at several PeV which is governed by
the abilities of hypernovae (HNe) accelerating protons and ULIRGs confining high energy protons.
However, the predicted PeV neutrino flux is so low that can not account for the IceCube data
unless the HNe rate in ULIRGs are $\sim $tens times higher than current estimate \cite{He2013}.
In the literature, the suggested sources for PeV neutrinos include active galactic neucleis (AGNs, \cite{Stecker2013}),
star forming galaxies (SFGs, \cite{Murase2013}), 
clusters of galaxies\cite{Murase2008}, hypernovae (HNe, \cite{He2013,Fox2013}),
gamma-ray bursts (GRBs, see \cite{Cholis2012}, but see \cite{He2012,Liu2012}),
and cosmogenic neutrinos (see \cite{Barger2012,Kalashev2013}, but see \cite{Bhattacharya2012,Roulet2012}).
In some literature, a single power-law spectrum of the TeV-PeV non-atmospheric neutrinos has been assumed (e.g., \cite{Anchordoqui2013,Laha2013}).    It is however not clear how robust such an assumption is since different astrophysical sources can produce neutrino emission in quite different energy ranges (e.g., \cite{Loeb2006,He2013,Murase2013,Stecker2013}) and the neutrino spectrum may display some wiggle-like structure. The main purpose of this work is to examine whether there is some tentative evidence for wiggle-like structure in current non-atmospheric neutrino data and then discuss the possible sources.

\begin{figure}
\includegraphics[width=95mm,angle=0]{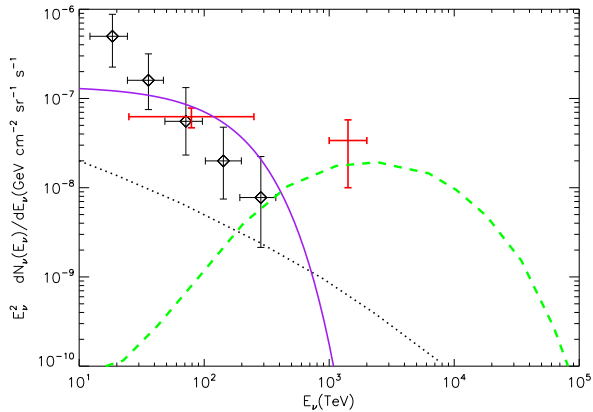}
\caption{A possible two-component spectrum model for the new IceCube data.
The convenience atmospheric muon neutrino emission is denoted by demand symbols,
 the prompt atmospheric muon neutrino emission is denoted by dotted line \cite{Whitehorn2013}
and the data for the non-atmospheric neutrinos observed by IceCube are denoted by plus symbols.
The solid line represents the lower energy component with a flat spectrum
followed by an exponential cutoff at the energy of $150$ TeV,
while the dashed line represents the neutrino spectrum produced via protons accelerated by AGNs interacting with the EBL
(adopted from \cite{Kalashev2013}).
}
\label{fig:model}
\end{figure}

To estimate the TeV-PeV non-atmospheric neutrino spectrum, Anchordoqui et al. \cite{Anchordoqui2013} partitioned the observation data into three energy bins, i.e., $50~{\rm TeV}-1$ PeV, $1-2$ PeV and $2-10$ PeV. Such a fit to the data suggests a single power-law spectrum index $\Gamma\sim 2.3$. We, however, notice that no events have been reliably detected between $250$ TeV and $1$ PeV,
but about 16 events observed over 10 atmospheric background neutrinos
in the energy range of $30-250$ TeV, and 2 events in the energy range of $1-2$ PeV \cite{Whitehorn2013}.
If the observed 18 non-atmospheric neutrinos have the same origin and can be described by a single power law spectrum
with index of $\Gamma$,  adopting the exposure of the full IceCube's 662 days operations
shown in \cite{Anchordoqui2013},
we can constrain the index via the flux in the two energy bins of $30-250$ TeV and $1-2$ PeV(The two data points are shown in Fig. \ref{fig:model}).
The fit to the data yields a $\Gamma\sim 2.2$, which suggests $\sim 9$
neutrinos in the energy range of $250$ TeV -- $1$ PeV,
while the data analysis found no reliable events in the same energy band. Such a fact favors the hypothesis that the TeV-PeV non-atmospheric neutrino spectrum may be not a single power-law, instead it may have structure.
For the single power law spectrum fits within $2\sigma$ error region,
 the lowest induced neutrino counts are 3.6.
Therefore, if the non-detection in the energy range of 250 TeV $-$ 1 PeV is intrinsic 
(i.e., neither due to the instrument effect nor limited by the data analysis method),
a single power-law spectrum model can be excluded at a confidence level of $\approx 2~\sigma$.
As a result, the TeV-PeV non-atmospheric neutrino spectrum may consist of two components, 
one has a cutoff at $\sim 250$ TeV and the other may have a sharp peak at $\sim 1$ PeV. 
We would like to caution that the above speculation is based on a small sample consisting of only $\sim 18$ events.
Since IceCube continues to collect data, our speculation will be directly confirmed or ruled out in the near future
. In the following discussion we focus on the two-component spectrum model and discuss the possible sources.

There are two fundamental processes to produce TeV-PeV neutrinos, i.e., the $pp$ collision and the $p\gamma$ interaction.
These two processes both produce charged pions and convert $20\%$ of the proton energy into pions.
The charged pions then  decay into neutrinos ($\pi^+\rightarrow \mu^++\nu_\mu\rightarrow e^++\nu_e+\bar\nu_\mu+\nu_\mu$
and $\pi^-\rightarrow \mu^-+\bar\nu_\mu\rightarrow e^-+\bar\nu_e+\nu_\mu+\bar\nu_\mu$).
Usually people assume that the decay products share the energy equally, then
$5\%$ of the proton energy will be converted into neutrinos.
Since the cross section of $pp$ collision changes very slowly with the energy of the ultra-relativistic protons,
the resulting neutrino spectrum traces that of the ultra-relativistic protons
and the cutoff energy $E_{\nu,\rm cut}$ of the neutrino spectrum mainly depends on the maximum energy $E_{p,\rm max}$ of accelerated protons,
i.e., $E_{\nu,\rm cut}\simeq 0.05E_{p,\rm max}$.
As a result of the $\Delta$ resonance,
the cross section of the $p\gamma$ interaction peaks at the photon energy $E'_{\gamma}\simeq 0.3$ GeV in the proton-rest frame.
Therefore, the resulting neutrino spectrum sensitively depends on not only the proton spectrum
but also the photon spectrum.
The neutrino spectrum will peak at energy of
$1.5\times10^{15}\Gamma^2/E_{\gamma,{\rm p}}$ eV,
where $E_{\gamma, {\rm p}}$ is the peak energy of photons in the observer's frame
and $\Gamma$ is the Lorentz factor of the emitting region\cite{He2012}.
Therefore, sources like SNe,
which can accelerate protons up to $5-10$ PeV in the observer's frame,
with dense circum medium surrounding, for example, in the SFGs,
may account for the $30-250$ TeV component via $pp$ collision.
And sources producing sufficient photons peaking at energy of $5$ eV,
i.e., around the ultraviolet (UV) band,
are possible to explain the PeV neutrino excess via $p\gamma$ process.
What's more, the thermal emission from the accretion disk of AGN and the EBL photons in UV band can be the target photons.
The observed peak energy of the target photons can be higher if considering the relativistic effect,
for instance, X-ray photons interacting with protons in a region with a Lorentz factor of tens
can also produce an observed neutrino spectrum peaking around PeV.

{\it The $30-250$ TeV component.}
SNe are widely suggested to be
the dominant sources for cosmic rays (CRs) at energies below the ``knee"
at $\sim 3\times 10^{15}\rm eV$, most probably through the diffusive shock
acceleration mechanism \cite{Hillas2005}.
The maximum energy of protons accelerated at the shock front of a SN expanding into the uniform dense interstellar
medium (ISM) can be estimated as \cite{Bell2001},
$\varepsilon_{\rm p,max}\approx 10^{16}{\rm eV}({V\over 3\times10^{8}~{\rm cm~s^{-1}}})^2
	({n\over 10^3\rm cm^{-3}})^{1/6}
	({M_{\rm SN}\over 10M_\odot})^{1/3}$,
where $n$ is the number density of ISM, and $M_{\rm SN}$ is the rest mass of the SN ejecta.
with a typical kinetic energy
$E_{\rm k}\sim 0.5-5\times 10^{51}$ erg and a typical velocity
$V \sim 3\times10^{8}~{\rm cm~s^{-1}}(E_{\rm k}/10^{51}~{\rm erg})^{1/2}(M_{\rm SN}/10M_\odot)^{-1/2}$.
The accelerated protons
would lose energy into $\gamma$-ray photons, electrons and positrons,
and neutrinos, through $pp$ collisions when injected into the ISM.

The energy loss time of protons is
$\tau_{\rm loss}=(0.5n\sigma_{\rm pp}c)^{-1}$,
where the factor $0.5$ is inelasticity \cite{Gaisser1990},
and $\sigma_{\rm pp}$ is the inelastic nuclear collision cross section,
which is $\sim 70 {\rm mb}$ for protons at energies
$\varepsilon'_{\rm p}\sim 1-10$ PeV in the rest frame that is of our great interest \cite{Alessandro2011}.
Introducing a parameter $\Sigma_{\rm gas}\equiv m_{\rm p}nl$
as the surface mass density of gas,
with $l$ as the scale of the dense region,
the energy loss time reads \cite{Condon1991}
\begin{equation}
\tau_{\rm loss}=1.4\times 10^{4}{\rm yr}\frac{l}{100\rm pc}\left(\frac{\Sigma_{\rm gas}}{1~{\rm g ~cm^{-2}}}\right)^{-1}.
\end{equation}
The confinement time can be rewritten as (i.e., Eq. (8) of \cite{He2013})
\begin{equation}
\tau_{\rm conf}\approx 2\times10^5{\rm yr}\left(\frac{\varepsilon'_{\rm p}}{10\rm PeV}\right)^{-0.5}(\frac{\Sigma_{\rm gas}}{1\rm g cm^{-2}})^{0.5}.
\end{equation}
The fraction of the energy that protons lose into pions is $f_\pi=1-\exp(-\tau_{\rm conf}/\tau_{\rm loss})$, which is close to 1 as long as
$\tau_{\rm conf}\geqslant\tau_{\rm loss}$.
As a result, the protons with energy $\varepsilon'_{\rm p}$
lose almost all of their energy via $pp$ collision
before escaping from the dense region
as long as $\tau_{\rm loss}\leq \tau_{\rm conf}$,
which constrains the critical gas surface density $\Sigma_{\rm crit}$ as
\begin{equation}
\Sigma_{\rm gas}\gtrsim\Sigma_{\rm crit}=0.17~{\rm g~cm^{-2}}
\left(\frac{\varepsilon'_{\rm p}}{10\rm PeV}\right)^{1/3}\left(\frac{l}{100\rm pc}\right)^{2/3}.
\label{eq:Sigma_gas}
\end{equation}
With the aim to explain neutrinos with energy of $30-250$ TeV,
we assume that the neutrinos spectrum cuts off at $\sim 200$ TeV.
To confine protons with the maximum energy of $4\times10^{15}$ eV,
the source should have gas surface density $\geq 0.13~{\rm g~cm^{-2}}$ if its typical scale is about $100$ pc.
Most starburst galaxies (SGs) satisfy such a request and hence are optimal candidates for accelerating protons and producing sub-PeV neutrinos.

As shown in \cite{He2013},
ULIRGs with the gas surface density about $1.0~{\rm g~cm^{-2}}$
can accelerate protons with energy up to 100 PeV by hypernovae (HNe),
confine the high energy protons and produce neutrinos with energy up to 5 PeV.
In contrast with ULIRGs,
the star formation rate (SFR) density of SGs
is $\sim 5$ times larger \cite{Magnelli2011, Magnelli2013}.
Consequently, the Type II SN rate is $\sim 5$ times larger than that estimated in \cite{He2013} since the Type II SN rate is proportional to
the SFR \cite{Fukugita2003}.
On the other hand, the ratio of the HN rate to the Type II SN rate is $\sim$ 0.01 and the typical ratio of the HN energy
to the Type II SN energy is $\sim 10$,
hence the energy budget of SGs for the PeV cosmic rays is  $\sim 50$ times larger
than that of ULIRGs for higher-energy protons.
As a result, the flux of sub-PeV neutrinos from SGs can be larger than that from ULIRGs by a factor of $\sim 50$,
suggesting a sub-PeV-neutrino flux from SGs $\sim 10^{-7}{\rm GeV\,cm^{-2}\, s^{-1}\, sr^{-1}}$,
consistent with the flux level of $6.25\pm 1.56 \times 10^{-8}{\rm GeV\, cm^{-2}\, s^{-1}\, sr^{-1}}$
observed by IceCube in $30-250$ TeV.

The neutrinos from SGs or SFGs have also been discussed in \cite{Loeb2006} and \cite{Murase2013}.
In \cite{Loeb2006}, these authors attributed the GHz radio emission of SGs to synchrotron emission
of secondary electrons produced via $pp$ collisions.
Then they normalized the flux of neutrinos at GeV via the total energy observed in GHz band being contributed by
GeV secondary electrons,
since the energy of protons converted into electrons and neutrinos are similar in the pion productions.
Depending on the spectral slopes, the extrapolated $<250$ TeV neutrino flux is in the range of $4\times10^{-9}-10^{-7}~{\rm GeV~ cm^{-2}~ s^{-1} ~sr^{-1}}$ (see Fig.1 of \cite{Loeb2006}),
which may be able to explain the new IceCube data, consistent with our result.
Murase et al. \cite{Murase2013} found out that SFGs 
can provide the necessary energy budget to explain the neutrino flux in PeV band,
but with the acceleration and confinement of $\sim 100$ PeV protons as the challenges.

{\it The PeV component.}
Stecker et al. \cite{Stecker1991} proposed a model to produce a neutrino spectrum
peaking at $1-10$ PeV,
via protons accelerated by shocks in the cores of AGNs
interacting with photons of the ``big blue bump" of
thermal emission from the accretion disk \cite{Laor1990}.
They estimated the value of the muon neutrino flux \cite{Stecker2007}
$\sim E_{\nu}^2\Phi(E_\nu)\sim 10^{-8} {\rm GeV cm^{-2} s^{-1} sr^{-1}}$ at $100$ TeV
and $\sim 6\times10^{-8} {\rm GeV cm^{-2} s^{-1} sr^{-1}}$ at $\sim 1$ PeV.
The uncertainty of the flux lies in the uncertainty of the AGN rate, the energy budget of AGNs
and the uncertainty of other model parameters, such as the spectrum of the accelerated protons.
So it's possible to explain the observed flux of $(3.28\pm2.28)\times10^{-8} {\rm GeV\,cm^{-2}\,s^{-1}\,sr^{-1}}$
in the energy range $1-2$ PeV \cite{Stecker2013}.

Another model with neutrino spectrum peaking at PeV is proposed in \cite{Kalashev2013},
which interprets the PeV neutrinos via
the EeV protons accelerated by AGNs
interacting with the EBL photons.
By adopting the EBL model with higher photons around $1{\mu m}$ 
and softer spectrum of accelerated protons with spectral index of $\alpha=2.6$,
they predict the flux of PeV neutrinos to be $(0.5-2)\times10^{-8}{\rm GeV\,cm^{-2}\,s^{-1}\,sr^{-1}}$,
consistent with the observed flux at PeV of $(3.28\pm2.28)\times10^{-8} {\rm GeV\,cm^{-2}\,s^{-1}\,sr^{-1}}$.
In Fig.\ref{fig:model}, we adopt such a scenario to account for the PeV excess.

{\it Summary and Discussions.}
Very recently the IceCube collaboration  has carried out the high-energy contained vertex search
to analysis the 662 days data from May 2010 to May 2012.
Such a search is sensitive at energies above $50$ TeV.
The main finding is 26 events with energy from $30$ TeV to $250$ TeV besides two $\sim 1$ PeV neutrinos,
while the atmospheric background is only expected to be $10.6^{+4.5}_{-3.9}$ events.
The TeV-PeV non-atmospheric neutrino detection has a confidence level of $4.3\sigma$ \cite{Whitehorn2013}.
The physical origin of the TeV-PeV non-atmospheric neutrino emission has been widely discussed in recent literature.
In this work we examine the possibility that the TeV-PeV non-atmospheric neutrino emission may originate from two groups of sources,
motivated by the non-detection of neutrinos in the energy range 250 TeV$-$ 1 PeV in current data.
If intrinsic, the non-detection of neutrinos in the energy range 250 TeV$-$ 1 PeV disfavors the single power-law spectrum model
for the TeV-PeV non-atmospheric neutrinos at a confidence level of $\sim 2\sigma$.
If the TeV-PeV non-atmospheric neutrino spectrum does consist two components, one likely has a flat spectrum
with a sharp cutoff at the energy $\sim 250$ TeV while the other peaks at $\sim 1$ PeV.
Interestingly, the former may be related to the Type II SNe which can accelerate protons to energies $\leq 10$ PeV
and these protons interact with the surrounding dense medium (via $pp$ collision), and produce charged pions which then decay into neutrinos.
The flat energy spectrum of the sub-PeV neutrinos can be straightforwardly understood
since the cross section of $pp$ collision changes very slowly with the energy of the ultra-relativistic protons.
We show that the starburst galaxies with high Type II SN rate and rich gas
are the promising sources of producing sub-PeV neutrinos.
The PeV neutrino excess component is hard to be interpreted within the $pp$ collision process.
Instead, the $p\gamma$ process can produce neutrino spectrum with a sharp peak
if the target photons have a very narrow energy distribution (e.g., \cite{Stecker1991,Kotera2010})
and has been adopted to interpret the PeV neutrino excess \cite{Kalashev2013}.

Our speculation that the TeV-PeV non-atmospheric neutrino emission may originate from two (or more) groups of sources
is based on a small sample consisting of only $\sim 18$ events. Much more data are needed to test whether
it is the case. Since IceCube continues to collect data, our speculation will be directly confirmed or ruled out in the near future.
If confirmed in the future, the wiggle-like structure of the TeV-PeV non-atmospheric neutrino spectrum
will shed valuable light on the underlying physical processes and the astrophysical sources.

%*****************************Fig.2***************************************
%*************************************************************************

{\it Acknowledgments.}  This work was supported in part by 973 Program of China under grant 2013CB837000,
National Natural Science of China under grants 11173064 and 11273063,
and by China Postdoctoral science foundation under grant 2012M521137 and 2013T60569,
and Jiangsu Province Postdoctoral science foundation under grant 1202052C.
YZF is also supported by the 100
Talents program of Chinese Academy of Sciences and the Foundation for
Distinguished Young Scholars of Jiangsu Province, China (No. BK2012047).

$^\ast$Corresponding author.\\
Electric addresses: hnhe@pmo.ac.cn,
bixian85@pmo.ac.cn,
yzfan@pmo.ac.cn,
 dmwei@pmo.ac.cn

\end{document}